\documentclass[preprint,12pt]{elsarticle}
\usepackage{amsmath}
\usepackage{amsfonts}
\usepackage{amssymb}
\usepackage[utf8]{inputenc}
\usepackage[T2A]{fontenc}
\usepackage{bm}

\begin{document}

\title{Bound electron states in a charged chain within the Dirac description} 
	
\author{Alexander Eremko, Larissa Brizhik, Vadim Loktev}
\address{Bogolyubov Institute for Theoretical Physics of the National Academy of Sciences of Ukraine \\
	Metrologichna Str., 14-b,  Kyiv, 03143, Ukraine.	\\[0pt]
	\vspace*{0.25cm}
	eremko@bitp.kyiv.ua, brizhik@bitp.kyiv.ua, vloktev@bitp.kyiv.ua }

\begin{abstract}
For the first time the exact analytical expressions for the three-dimensional bound electron states in the Coulomb field of the chain consisting  of positively charged ions, are obtained within the Dirac description, using the new spinor invariant found for this problem. It is demonstrated that within such approach the coupling between electron spin and its one-dimensional propagation along the chain  naturally arise, without any need to include artificially into the equations the so-called spin-orbit interaction.

\end{abstract}

\vspace{10mm}

\maketitle

\section{Introduction}\label{Intro} 

It is well known that soliton concept is widely used in description of numerous physical processes, among which there are hydrodynamic (solitary wave), elastic (nonlinear sound waves, dislocations), optical (solitons in optical waveguides), electroconductivity (electrosolitons), magnetic (magnetic solitons, domain walls, etc.), plasma systems, and many other (see, e.g., \cite{DauPeyr,KosIvKov}). The pioneering work of A.S. Davydov and N.I. Kislukha \cite{DavydovKislukha1} were the first to introduce the concept of solitons in condensed matter physics. Davydov's soliton \cite{Davydov,Scott} is to a large extent especial for it exists in one-dimensional (1D) systems. It represents an  electron self-trapped in the deformational potential of the system  due to electron-lattice interaction  \cite{Scott,ground} (see also \cite{KovalevKosevich}). To-day the concept of the Davydov's soliton  is used to explain energy and charge transfer in biological systems  \cite{DavQBiol,DauPeyr}, charge transport on macroscopic distances in Donor -- Quantum Wire -- Acceptor systems  \cite{Brizhik-DAA} and many other related problems. 

Theory of the Davydov's soliton was based on the assumption that electron states in 1D structures can be described within the tight binding model using non-relativistic Shr\"odinger equation (SE). In the meantime it has been shown in our recent papers  \cite{BEL1,BEL2}, that within the Dirac equation (DE) the number of possible solutions is enlarged with the possess unexpected properties of the new solutions. This has been done for essentially linear problem and self-trapping phenomenon has not been considered. Its possibility and criteria of existence can be studied when electron eigenstates, and, therefore, eigenfunctions of the equation with the potential of the considered structure  are known. 

In the present paper, dedicated to Alexander Kovalev, one of the best experts in the field of theory of magnetic solitons in condensed matter, we aim to study self-trapped Dirac states of an electron in a chain which so far to our knowledge are unknown. In particular, here we solve only the first part of the nonlinear problem of  soliton formation within the Dirac description. For this we calculate analytically eigenstates and eigenfunctions of an electron in a three-dimensional (3D) field of a charged chain, using the set of the invariants of the DE. Each invariant leads to the corresponding solution.  They can be entangled, according to the general invariant as a linear combination of the three initial ones.

\section[Model]{Description of the model} 

Let us consider an atomic chain with regularly placed $ N \gg 1 $ atoms which is described by the radius-vector  $ \mathbf{r}_{n} = \lbrace 0, 0, na \rbrace $ ($ n = 1,2,\ldots \,, N $). Each atom has the charge $ Q $ and, therefore, such a chain creates for electrons the potential 
\begin{equation}
	\label{pot}
V\left( \mathbf{r} \right) = - \sum_{n} \frac{eQ}{\sqrt{x^{2} + y^{2} + (z - na)^{2}}}  
\end{equation}
where the radius-vector can be written in the form
\begin{equation}
	\label{rad-v}
\mathbf{r}=\left( \mathbf{r}_{\perp},z \right), \quad \mathbf{r}_{\perp}^2=x^2+y^2.	
\end{equation}
This potential has the translational symmetry along the chain $ V\left( \mathbf{r} + na\mathbf{e}_{z} \right) = V\left( \mathbf{r} \right) $ and, as it is well known, can be represented in the form 
\begin{equation}
\label{V(r)}  
V\left( \mathbf{r} \right) = \sum_{g_{n}} a_{g_{n}} \left( \mathbf{r}_{\perp} \right) e^{ig_{n}z}
\end{equation}
where $ g_{n} = 2\pi n/a $ with $ 2\pi/a $ is a reciprocal length of 1D lattice with the lattice spacing $ a $. The coefficients $ a_{g_{n}} \left( \mathbf{r}_{\perp} \right) $  are given by the expression 
\begin{equation}
\label{a_g} 
a_{g_{n}} \left( \mathbf{r}_{\perp} \right) = \frac{1}{a} \int_{(\text{over cell})} V\left( \mathbf{r} \right) e^{-ig_{n}z} dz = -\frac{eQ}{a} \int_{-a/2}^{a/2} \frac{e^{-ig_{n}z}}{\sqrt{\mathbf{r}_{\perp}^{2} + z^{2}}} dz . 
\end{equation} 
Here the Wigner-Seitz cell with an atom in its center is chosen as an elementary cell of 1D Bravais lattice. 

The most adequate and complete description of the electron states is provided by the  (DE) which in the presence of an external field created by atomic nuclei, is 
\begin{equation}
\label{DE} 
\left( c \bm{\hat{\alpha}} \cdot \hat{\mathbf{p}} + m c^{2} \hat{\beta} + V \left(\mathbf{r} \right) \right) \Psi = E \Psi  
\end{equation} 
where $ c \bm{\hat{\alpha}} \cdot \hat{\mathbf{p}} + m c^{2} \hat{\beta} + V \left(\mathbf{r} \right) = \hat{H}_{D} $ is Dirac Hamiltonian, $ \Psi = \Psi\left(\mathbf{r} \right) $ is Dirac bispinor, and $ V \left(\mathbf{r} \right) = \sum_{g_{n}} V_{g_{n}} \left( \mathbf{r}_{\perp},z \right) $ is the  potential created by the atomic chain, determined in Eq. (\ref{V(r)}).  Matrices $ \hat{\beta} $ and components $ \hat{\alpha}_{j} $ ($ j=x,y,z $) of the vector-matrix $ \bm{\hat{\alpha}} = \sum_{j} \mathbf{e}_{j} \hat{\alpha}_{j} $ together with the unit matrix  $ \hat{\mathrm{I}} $ are $ 4\times 4 $ Hermitian Dirac matrices (for more explanation see Appendix). The Fourier coefficients $ a_{g_{n}} $, in (\ref{V(r)}), and, therefore, $ V_{g_{n}} $ decrease with increasing $ g_{n} $. 

\section{Free quasi-one-dimensional electrons} 

As the zero-order  approximation we will use the solution of the DE (\ref{DE}) with the potential $ V ( \mathbf{r}) $ given by zero-order Fourier component $ V_{0} = V_{g=0} $, only, 
\[
V_{0} \left( \mathbf{r}_{\perp} \right) = -\frac{eQ}{a} \int_{-a/2}^{a/2} \frac{dz}{\sqrt{\mathbf{r}_{\perp}^{2} + z^{2}}} = \frac{eQ}{a} \ln \dfrac{\sqrt{\mathbf{r}_{\perp}^{2} + \frac{1}{4}a^{2}} - \frac{1}{2}a}{\sqrt{\mathbf{r}_{\perp}^{2} + \frac{1}{4}a^{2}} + \frac{1}{2}a} = \frac{2eQ}{a} \ln \dfrac{r_{\perp}}{\sqrt{r_{\perp}^{2} + \frac{1}{4}a^{2}} + \frac{1}{2}a} ,
\] 
which we write down as 
\begin{equation}
\label{V_0} 
V_{0} \left( \mathbf{r}_{\perp} \right) = - \frac{2eQ}{a} \ln v\left( r_{\perp} \right) , \quad 
v\left( r_{\perp} \right) = \frac{a}{2r_{\perp}} + \sqrt{1 + \frac{a^{2}}{4r_{\perp}^{2}}} . \quad r_{\perp} = |\mathbf{r}_{\perp}| = \sqrt{x^{2} + y^{2}} . 
\end{equation} 

The Dirac Hamiltonian $ H_{D} $ is presented through Dirac $ 4\times 4 $ matrices and is often written down in the $ 2\times 2 $ block form 
\begin{equation}
\label{H_D-m} 
\hat{H}_{D} = \left( \begin{array}{cc}
\left( V_{0}+mc^{2} \right) \hat{\mathbb{I}}_{2} & c\bm{\hat{\sigma}} \cdot \mathbf{\hat{p}} \\
 c\bm{\hat{\sigma}} \cdot \mathbf{\hat{p}} & \left( V_{0}-mc^{2} \right) \hat{\mathbb{I}}_{2}
\end{array}  \right).
\end{equation}
Here $ \hat{\mathbb{I}}_{2} $ is a unit $ 2\times 2 $ matrix, $ \bm{\hat{\sigma}} = \sum_{j} \mathbf{e}_{j} \hat{\sigma}_{j} $ where $ \hat{\sigma}_{j} $ ($ j=x,y,z $) are Pauli matrices. Respectively, the bispinor $ \Psi $ is represented as
\begin{equation}
\label{bispinor} 
\Psi \left( \mathbf{r} \right) = \left( \begin{array}{c} \psi^{(u)} \left( \mathbf{r} \right)  \\ \psi^{(d)} \left( \mathbf{r} \right) 
\end{array} \right) , \quad 
\psi^{(u)} = \left( \begin{array}{c} 
\psi^{(1)} \left( \mathbf{r} \right)  \\ \psi^{(2)} \left( \mathbf{r} \right) 
\end{array} \right) , \; \psi^{(d)} = \left( \begin{array}{c}
\psi^{(3)} \left( \mathbf{r} \right)  \\ \psi^{(4)} \left( \mathbf{r} \right) 
\end{array} \right) , 
\end{equation}
where $ \psi^{(u/d)} $ are its upper/lower spinors, respectively, with the components $ \psi^{(\nu )} $ and $ \psi^{(\nu + 1)} $  ($ \nu =1 $ for the upper bispinor, and $ \nu = 3 $ for the lower one). 

Therefore, the  DE (\ref{DE}) is the system of two equations for upper and lower spinor components $ \psi^{(u/d)} $ of the bispinor (\ref{bispinor})
\begin{equation}
\label{DsysEq_1} 
\begin{array}{c} 
c\bm{\hat{\sigma}} \cdot \mathbf{\hat{p}} \psi^{(u)} - 
\left( E  - V_{0}(r_{\perp}) \right) \psi^{(d)} - mc^{2} \psi^{(d)} = 0 , \\ 
c\bm{\hat{\sigma}} \cdot \mathbf{\hat{p}} \psi^{(d)} - 
\left( E - V_{0} (r_{\perp}) \right) \psi^{(u)} + mc^{2} \psi^{(u)} = 0  .
\end{array}
\end{equation} 

Solutions of the DE system (\ref{DsysEq_1}) determine stationary states of quasi-1D electrons bound by an atomic chain. Stationary states are characterized by the set of quantum numbers which reflect the eigenvalues of the  complete set of operators that commute with the Hamiltonian $ \hat{H}_{D} $ (constants of motion or invariants). In the potential $ V_{0}\left( \mathbf{r}_{\perp} \right) $ with the translational symmetry along $ z $-axis, such invariants are $ z $-components of the momentum $ \hat{p}_{z} $, of the total angular momentum $ \hat{J}_{z} = \hat{L}_{z} \hat{\mathbb{I}} + (\hbar/2) \hat{\Sigma}_{z} $, and of the spin polarization vector $ \hat{\mathcal{S}}_{z} = \hat{\Omega}_{z} + \hat{\rho}_{1} \frac{\hat{p}_{z}}{mc} $. Here the total angular momentum is $ \mathbf{\hat{J}} = \hat{\mathbf{L}} + (\hbar /2) \bm{\hat{\Sigma}} $ where $ \hat{\mathbf{L}} = \mathbf{r}\times \mathbf{\hat{p}} $ is the orbital momentum operator, and $ \left(\hbar /2 \right) \bm{\hat{\Sigma}} $ is the operator of the spin angular momentum, called "spin" for short.  

Thus, the eigen bispinors of the DE are the joint eigenstate vectors of the complete set of the independent commuting operators (including the Hamiltonian) and the set of quantum numbers is determined by the eigenvalue equations 
\begin{equation} 
	\label{eq1} 
\begin{array}{c}
\hat{H}_{D} \Psi_{E,k,M,s} = E \Psi_{E,k,M,s} , \quad \hat{p}_{z} \Psi_{E,k,M,s} = \hbar k \Psi_{E,k,M,s} , \\
\hat{J}_{z} \Psi_{E,k,M,s} = \hbar M \Psi_{E,k,M,s} , \quad \hat{\mathcal{S}}_{z} \Psi_{E,k,M,s} = s \Psi_{E,k,M,s}  
\end{array}  
\end{equation}
where $k\equiv k_{z} $ is the wave number along the chain, $ M$ is the eigenvalue of $J_z$.

\subsection{Eigenvalues and eigen bispinors of the invariants}

It follows from Eq. (\ref{eq1}) that $ \Psi_{E,k,M,s} = e^{ikz}\Psi_{E,k,M,s}(x,y) $.  Due to the symmetry of $ V_{0} $ it is naturally to use the cylindrical coordinate system 
\[ 
x = r_{\perp} \cos \varphi ,  \quad y = r_{\perp} \sin \varphi , \quad z . 
\] 

The eigenvalue equation for $ z $-component of the total angular momentum $ \hat{J}_{z} $ is
\[ 
\left( \begin{array}{cc}
\hat{j}_{z} & 0 \\ 0 & \hat{j}_{z}
\end{array}  \right) \left( \begin{array}{c} \psi^{(u)} \\ \psi^{(d)} \end{array} \right) = \hbar M \left( \begin{array}{c} \psi^{(u)} \\ \psi^{(d)} \end{array} \right) , \quad 
\hat{j}_{z} = \hat{L}_{z} \hat{\mathbb{I}}_{2} + \frac{\hbar}{2} \hat{\sigma}_{z} = \left( \begin{array}{cc} 
\hat{L}_{z} + \frac{\hbar}{2} & 0 \\
0 & \hat{L}_{z} - \frac{\hbar}{2} 
\end{array} \right) 
\] 
where $ \hat{L}_{z} $ is $ z $-component of the orbital momentum,or, equivalently,  $ \hat{j}_{z} \psi^{(u/d)} = \hbar M \psi^{(u/d)} $ and, according to the latter equation, the upper and lower spinors are the eigen spinors of $ \hat{j}_{z} $. In the cylindrical coordinates $ \hat{L}_{z} = -i\hbar \partial /\partial \varphi $ and we have two equations 
\[
-i\hbar \frac{\partial \psi^{(\nu)}}{\partial \varphi} + \frac{\hbar}{2} \psi^{(\nu)} = \hbar M \psi^{(\nu)} , \quad -i\hbar \frac{\partial \psi^{(\nu + 1)}}{\partial \varphi} - \frac{\hbar}{2} \psi^{(\nu + 1)} = \hbar M \psi^{(\nu + 1)} 
\]
for the bispinor components $ \psi^{(\nu)} (\mathbf{r}_{\perp},z,\varphi) $ with $ \nu = 1 $ for the upper spinor $ \psi^{(u)} $, and $ \nu = 3 $ for the lower one, $ \psi^{(d)} $, in (\ref{bispinor}). Solutions of these equations are 
\[
\psi^{(\nu)} \sim e^{im_{1}\varphi} \quad \text{and} \quad \psi^{(\nu + 1)} \sim e^{im_{2}\varphi}
\]
where $ m_{1} $ and $ m_{2} $ are the integer numbers that satisfy the condition $ m_{2} - m_{1} = 1 $. Thus, the bispinor components are 
\begin{equation}
\label{comps1} 
\psi_{k,M}^{(u/d)} = e^{i \left( kz + M\varphi \right) } \chi^{(u/d)} , \quad 
\chi^{(u)} = \left( \begin{array}{c} 
e^{-i\varphi /2} f_{1} (\mathbf{r}_{\perp}) \\ 
e^{i\varphi /2} f_{2} (\mathbf{r}_{\perp}) \end{array} \right) , \quad 
\chi^{(d)} = \left( \begin{array}{c} 
e^{-i\varphi /2} f_{3} (\mathbf{r}_{\perp}) \\ 
e^{i\varphi /2} f_{4} (\mathbf{r}_{\perp}) \end{array} \right) 
\end{equation}
with quantum numbers $ k $ (recall,  it is the wave number along the chain), and numbers $ M = (1/2) (m_{2} + m_{1}) $ which take half-integer values $ M = \pm 1/2, \pm 3/2 , \ldots $. 

The eigenvalue equation for $ z $-component of the spin polarization vector $ \hat{\mathcal{S}}_{z} $ is 
\[
\left( \begin{array}{cc}
\hat{\sigma}_{z}  & \hat{\mathbb{I}}_{2} \frac{\hat{p}_{z}}{mc} \\ \hat{\mathbb{I}}_{2} \frac{\hat{p}_{z}}{mc} & - \hat{\sigma}_{z},
\end{array}  \right) \left( \begin{array}{c} \psi^{(u)} \\ \psi^{(d)} \end{array} \right) = s \left( \begin{array}{c} \psi^{(u)} \\ \psi^{(d)} \end{array} \right) 
\] 
which, with account of Eq. (\ref{comps1}), leads to the equations 
\begin{equation}
\label{EqsS_z} 
\begin{array}{cc}
\left( s - \hat{\sigma}_{z} \right) \psi_{k,M}^{(u)} = \frac{\hbar k}{mc} \psi_{k,M}^{(d)} , \\ 
\left( s + \hat{\sigma}_{z} \right) \psi_{k,M}^{(d)} = \frac{\hbar k}{mc} \psi_{k,M}^{(u)} . 
\end{array}
\end{equation}
Multiplying the first equation by $ \left( s + \hat{\sigma}_{z} \right) $ and the second one by $ \left( s - \hat{\sigma}_{z} \right) $, we get the relation
\[
\left( s^{2} - 1 - \frac{\hbar^{2} k^{2}}{m^{2}c^{2}} \right) \psi_{k,M}^{(u/d)} = 0
\]
from which we obtain two eigenvalues
\begin{equation}
\label{eignvalS} 
s_{\pm} = \pm \sqrt{1 + \frac{\hbar^{2}k^{2}}{m^{2}c^{2}}} = \sigma s 
\end{equation} 
 of the operator $ \hat{\mathcal{S}}_{z} $  where  notations are introduced  
\begin{equation}
\label{calE_k} 
\sigma = \pm 1, \quad s = \sqrt{1 + \frac{\hbar^{2}k^{2}}{m^{2}c^{2}}} = \frac{\mathcal{E}_{k}}{mc^{2}} , \quad \mathcal{E}_{k} = \sqrt{m^{2}c^{4} + c^{2} \hbar^{2} k^{2}}.  
\end{equation}

The last eigenvalue equation gives us the two-valued spin quantum number $ \sigma = \pm 1 $ and, respectively, two solutions $ \Psi_{E,k,M,\sigma} $ with opposite signs of $ \sigma $. According to Eq. (\ref{EqsS_z}), the upper and lower spinors for each spin state are connected by the relation which can be written in the two equivalent forms
\begin{equation}
\label{rel1S} 
\left( \sigma \mathcal{E}_{k} + mc^{2} \hat{\sigma}_{z} \right) \psi_{k,M,\sigma}^{(d)} = c\hbar k \psi_{k,M,\sigma}^{(u)} , \quad 
\left( \sigma \mathcal{E}_{k} - mc^{2} \hat{\sigma}_{z} \right) \psi_{k,M,\sigma}^{(u)} = c\hbar k \psi_{k,M,\sigma}^{(d)} ,  
\end{equation} 
because $ \left( \sigma \mathcal{E}_{k} + mc^{2} \hat{\sigma}_{z} \right) \left( \sigma \mathcal{E}_{k} - mc^{2} \hat{\sigma}_{z} \right) = \mathcal{E}_{k}^{2} - m^{2}c^{4} = c^{2} \hbar^{2} k^{2} $.

Adding and subtracting these two equalities, we get 
\begin{equation}
\label{relS+,-} 
mc^{2} \hat{\sigma}_{z} \psi_{k,M,\sigma}^{(-)} = \left( \sigma \mathcal{E}_{k} - c\hbar k \right) \psi_{k,M,\sigma}^{(+)} , \quad 
mc^{2} \hat{\sigma}_{z} \psi_{k,M,\sigma}^{(+)} = \left( \sigma \mathcal{E}_{k} + c\hbar k \right) \psi_{k,M,\sigma}^{(-)}
\end{equation}
where $ \psi_{k,M,\sigma}^{(\pm)} = \psi_{k,M,\sigma}^{(u)} \pm \psi_{k,M,\sigma}^{(d)} $  . 

\subsection{Solution of the Dirac equation} 

To find the solution of the DE, we take into account the existence of the constants of motion and  use the spinors determined in Eq. (\ref{comps1}) in the DE system (\ref{DsysEq_1}). Because $ \hat{p}_{z} $ is the constant of motion, we represent the momentum operator in Eqs. (\ref{DsysEq_1}) as $ \hat{\mathbf{p}} = \hat{\mathbf{p}}_{\perp} + \hat{p}_{z} \mathbf{e}_{z} $ where $ \hat{\mathbf{p}}_{\perp} = \hat{p}_{x} \mathbf{e}_{x} + \hat{p}_{y} \mathbf{e}_{y} $. By adding and subtracting two spinor equations in (\ref{DsysEq_1}) we transform it to the system of equations for spinors $ \psi_{k,M,\sigma}^{(\pm)} $  
\[
\begin{array}{c} 
c\bm{\hat{\sigma}} \cdot \mathbf{\hat{p}}_{\perp} \psi_{k,M,\sigma}^{(+)} + c \hat{\sigma}_{z} \hat{p}_{z} \psi_{k,M,\sigma}^{(+)} - \left( E - V_{0}(r_{\perp}) \right) \psi^{(+)} + mc^{2} \psi_{k,M,\sigma}^{(-)} = 0 , \\ 
c\bm{\hat{\sigma}} \cdot \mathbf{\hat{p}}_{\perp} \psi_{k,M,\sigma}^{(-)} + c \hat{\sigma}_{z} \hat{p}_{z} \psi_{k,M,\sigma}^{(-)} + \left( E - V_{0} (\mathbf{r}_{\perp}) \right) \psi_{k,M,\sigma}^{(-)} - mc^{2} \psi_{k,M,\sigma}^{(+)} = 0 .
\end{array}
\]
where according to (\ref{comps1}), spinors $ \psi_{k,M,\sigma}^{(\pm)} $ are given by the formula
\begin{equation}
\label{spinor+,-} 
\psi_{k,M,\sigma}^{(\pm)} = e^{i \left( kz + M\varphi \right) } \chi_{\sigma}^{(\pm)} 
\end{equation}
in which the notations are used
\begin{equation}
	\label{chi-s}
\chi_{\sigma}^{(\pm)} = \left( \begin{array}{c} 
	e^{-i\varphi /2} f_{\sigma}^{(\pm)} (\mathbf{r}_{\perp}) \\ 
	e^{i\varphi /2} g_{\sigma}^{(\pm)} (\mathbf{r}_{\perp}) \end{array} \right) ,  \quad 
\begin{array}{c} 
	f_{\sigma}^{(\pm)} (\mathbf{r}_{\perp}) = f_{1} (\mathbf{r}_{\perp}) \pm f_{3} (\mathbf{r}_{\perp}) \\ 
	g_{\sigma}^{(\pm)}(\mathbf{r}_{\perp}) = f_{2} (\mathbf{r}_{\perp}) \pm f_{4} (\mathbf{r}_{\perp}) \end{array}  
\end{equation}
 
To write down equations in cylindrical coordinate system, we use the transformation of the kinetic energy operator $ \bm{\hat{\sigma}} \mathbf{\hat{p}}_{\perp} $. For this we define $ \bm{\mathbf{r}_{\perp}} = x \mathbf{e}_{x} + y \mathbf{e}_{y} $, $ \bm{\mathbf{r}_{\perp}}^{2} = x^{2} + y^{2} = r_{\perp}^{2} $ and the unit matrix  
\begin{equation} 
\label{sigma_rho} 
\hat{\sigma}_{\mathbf{r}_{\perp}} = \bm{\hat{\sigma}} \cdot \mathbf{e}_{\bm{\mathbf{r}_{\perp}}} = \left( 
\begin{array}{cc}
0 & e^{-i\varphi} \\ e^{i\varphi} & 0
\end{array} \right) , \quad \mathbf{e}_{\bm{\mathbf{r}_{\perp}}} = \frac{\bm{\mathbf{r}_{\perp}}}{r_{\perp}} , \quad \hat{\sigma}_{\mathbf{r}_{\perp}}^{2} = \hat{\mathbb{I}}_{2},
\end{equation} 
and take into account the identity 
$$ 
\hat{\sigma}_{\mathbf{r}_{\perp}} \left( \bm{\hat{\sigma}} \cdot \mathbf{\hat{p}}_{\perp} \right) = \frac{1}{\mathbf{r}_{\perp}} \left( \bm{\hat{\sigma}}\cdot \bm{\mathbf{r}_{\perp}} \right) \left( \bm{\hat{\sigma}} \cdot \mathbf{\hat{p}} \right) = \frac{1}{r_{\perp}} \left( \bm{\mathbf{r}_{\perp}} \cdot \mathbf{\hat{p}} + i \bm{\hat{\sigma}} \cdot \bm{\mathbf{r}_{\perp}} \times \mathbf{\hat{p}} \right) = \hat{p}_{r_{\perp}} + \frac{i}{r_{\perp}} \hat{\sigma}_{z} \hat{j}_{z} .
$$ 
Here $ \hat{p}_{r_{\perp}} = (1/2)\left( \mathbf{e}_{\bm{\mathbf{r}_{\perp}}} \cdot \mathbf{p}_{\perp} + \mathbf{p}_{\perp} \cdot \mathbf{e}_{\bm{\mathbf{r}_{\perp}}} \right)  $ is the Hermitian operator of the momentum projection on the direction $ \bm{\mathbf{r}_{\perp}} $ which in cylindrical coordinates is 
\begin{equation}
\label{p_rho} 
\hat{p}_{r_{\perp}} = - i\hbar \left( \frac{\partial}{\partial r_{\perp}} + \frac{1}{2r_{\perp}} \right) = - i\hbar \frac{1}{\sqrt{r_{\perp}}} \frac{\partial}{\partial r_{\perp}} \sqrt{r_{\perp}} , 
\end{equation} 
and $ \hat{j}_{z} $ is a diagonal block of $ \hat{J}_{z} $. 

So, the operator of the kinetic energy of the electron motion  transverse to the chain, is represented as 
$$ c\bm{\hat{\sigma}} \cdot \mathbf{\hat{p}}_{\perp} = c \hat{\sigma}_{\mathbf{r}_{\perp}}^{2}\bm{\hat{\sigma}} \cdot \mathbf{\hat{p}}_{\perp} = 
\hat{\sigma}_{\mathbf{r}_{\perp}} c \left( \hat{p}_{r_{\perp}} + \frac{i}{r_{\perp}} \hat{\sigma}_{z} \hat{j}_{z} \right) . $$ 

 Taking into account that $ \hat{\sigma}_{\mathbf{r}_{\perp}}^{2} = \hat{\mathbb{I}}_{2} $ and the explicit expressions (\ref{spinor+,-}), equations for the spinors $ \psi_{\sigma}^{(\pm)} $ become 
\[
\begin{array}{c} 
c \hat{p}_{r_{\perp}} \chi_{\sigma}^{(+)} + i \frac{c\hbar M}{r_{\perp}} \hat{\sigma}_{z} \chi_{\sigma}^{(+)} 
+ c\hbar k \hat{\sigma}_{\mathbf{r}_{\perp}} \hat{\sigma}_{z} \chi_{\sigma}^{(+)} - \left( E - V_{0}(r_{\perp}) \right) \hat{\sigma}_{\mathbf{r}_{\perp}} \chi_{\sigma}^{(+)} + mc^{2} \hat{\sigma}_{\mathbf{r}_{\perp}} \chi_{\sigma}^{(-)} = 0 ,   
 \\ 
c \hat{p}_{r_{\perp}} \chi_{\sigma}^{(-)} + i \frac{c\hbar M}{r_{\perp}} \hat{\sigma}_{z} \chi_{\sigma}^{(-)} 
+ c\hbar k \hat{\sigma}_{\mathbf{r}_{\perp}} \hat{\sigma}_{z} \chi_{\sigma}^{(-)}  + \left( E - V_{0} (r_{\perp}) \right) \hat{\sigma}_{\mathbf{r}_{\perp}} \chi_{\sigma}^{(-)} - mc^{2} \hat{\sigma}_{\mathbf{r}_{\perp}} \chi_{\sigma}^{(+)} = 0 .
\end{array}
\] 
It follows from relations (\ref{relS+,-}) that
\[
\chi_{\sigma}^{(-)} = \frac{mc^{2}}{\sigma \mathcal{E}_{k} + c\hbar k } \hat{\sigma}_{z} \chi_{\sigma}^{(+)} , \quad \chi_{\sigma}^{(+)} = \frac{mc^{2}}{\sigma \mathcal{E}_{k} - c\hbar k } \hat{\sigma}_{z} \chi_{\sigma}^{(-)} .
\]
Next, we introduce from the relation $ \hat{\sigma}_{z} \hat{\sigma}_{\mathbf{r}_{\perp}} = i\hat{\sigma}_{\varphi} $ the unit matrix 
\[ 
\hat{\sigma}_{\varphi} =  \left( \begin{array}{cc}
0 & -i e^{-i\varphi} \\ i e^{i\varphi} & 0
\end{array} \right) .
\]
In view of the equality $ \hat{p}_{r_{\perp}}\chi = (-i\hbar /\sqrt{r_{\perp}}) \partial \sqrt{r_{\perp}}\chi)/\partial r_{\perp} $, it is convenient to consider spinors $ \Phi_{\sigma}^{(\pm)} = \sqrt{r_{\perp}}\chi_{\sigma}^{(\pm)} $ using substitution $ f_{\sigma}^{(\pm)}(r_{\perp}) = r_{\perp}^{-1/2} F_{\sigma}^{(\pm)}(r_{\perp}) $ and $ g_{\sigma}^{(\pm)}(r_{\perp}) = r_{\perp}^{-1/2} G_{\sigma}^{(\pm)}(r_{\perp}) $ in spinors (\ref{spinor+,-}). This leads to two independent equations for the new spinors $ \Phi_{\sigma}^{(\pm)} $ 
\[
\begin{array}{c} 
- ic\hbar \frac{\partial}{\partial r_{\perp}} \Phi_{\sigma}^{(+)} + i \frac{c\hbar M}{r_{\perp}} \hat{\sigma}_{z} \Phi_{\sigma}^{(+)} - ic\hbar k \hat{\sigma}_{\varphi} \Phi_{\sigma}^{(+)} 
- \left( E - V_{0} \right) \hat{\sigma}_{\mathbf{r}_{\perp}} \Phi_{\sigma}^{(+)} - i\frac{m^{2}c^{4}}{\sigma \mathcal{E}_{k} + c\hbar k} \hat{\sigma}_{\varphi} \Phi_{\sigma}^{(+)} = 0  , \\ 
- ic\hbar \frac{\partial}{\partial r_{\perp}} \Phi_{\sigma}^{(-)} + i \frac{c\hbar M}{r_{\perp}} \hat{\sigma}_{z} \Phi_{\sigma}^{(-)} - ic\hbar k \hat{\sigma}_{\varphi} \Phi_{\sigma}^{(-)} 
+ \left( E - V_{0} \right) \hat{\sigma}_{\mathbf{r}_{\perp}} \Phi_{\sigma}^{(-)} + i\frac{m^{2}c^{4}}{\sigma \mathcal{E}_{k} - c\hbar k} \hat{\sigma}_{\varphi} \Phi_{\sigma}^{(-)} = 0  .
\end{array}
\] 

Action of the Pauli matrices $ \hat{\sigma}_{z} $, $ \hat{\sigma}_{\varphi} $, and $ \hat{\sigma}_{\mathbf{r}_{\perp}} $ on these spinors gives the following equations:
\[
\hat{\sigma}_{z} \Phi = 
 \left( \begin{array}{c} 
e^{-i\varphi /2} F (r_{\perp}) \\ 
- e^{i\varphi /2} G (r_{\perp}) \end{array} \right) , \quad 
\hat{\sigma}_{\varphi} \Phi = 
= i \left( \begin{array}{c} 
	- e^{-i\varphi /2} G (r_{\perp}) \\ 
	e^{i\varphi /2} F (r_{\perp}) \end{array} \right), \quad 
\hat{\sigma}_{\bm{\mathbf{r}_{\perp}} } \Phi 
=\left( \begin{array}{c} 
	e^{-i\varphi /2} G (r_{\perp}) \\ 
	e^{i\varphi /2} F (r_{\perp}) \end{array} \right)	
\] 
where for simplicity we have omitted upper and lower indices. Respectively, the spinor equations can be represented as the systems of two sets of equations
\[
\begin{array}{c}
-ic\hbar \frac{dF_{\sigma}^{(+)}}{dr_{\perp}} + i\frac{c\hbar M}{\mathbf{r}_{\perp}} F_{\sigma}^{(+)} - \left( E + \sigma \mathcal{E}_{k} \right) G_{\sigma}^{(+)} + V_{0}(r_{\perp}) G_{\sigma}^{(+)} = 0 , \\ 
-ic\hbar \frac{dG_{\sigma}^{(+)}}{dr_{\perp}} - i\frac{c\hbar M}{\mathbf{r}_{\perp}} G_{\sigma}^{(+)} - \left( E - \sigma \mathcal{E}_{k} \right) F_{\sigma}^{(+)} + V_{0}(r_{\perp}) F_{\sigma}^{(+)} = 0 ,
\end{array}
\]
and 
\[
\begin{array}{c}
-ic\hbar \frac{dF_{\sigma}^{(-)}}{dr_{\perp}} + i\frac{c\hbar M}{\mathbf{r}_{\perp}} F_{\sigma}^{(-)} + \left( E + \sigma \mathcal{E}_{k} \right) G_{\sigma}^{(-)} - V_{0}(r_{\perp}) G_{\sigma}^{(-)} = 0 , \\ 
-ic\hbar \frac{dG_{\sigma}^{(-)}}{dr_{\perp}} - i\frac{c\hbar M}{\mathbf{r}_{\perp}} G_{\sigma}^{(-)} + \left( E - \sigma \mathcal{E}_{k} \right) F_{\sigma}^{(-)} - V_{0}(r_{\perp}) F_{\sigma}^{(-)} = 0 ,
\end{array}
\] 
which shows that the DE equation (\ref{DsysEq_1}) is the system of four equations which determine four components of the Dirac bispinor $ \Psi_{E,k,M,\sigma} $. Here we have used again the relation $ \left( \sigma \mathcal{E}_{k} + c\hbar k \right) \left( \sigma \mathcal{E}_{k} - c\hbar k \right) = \mathcal{E}_{k}^{2} - c^{2}\hbar^{2} k^{2} = m^{2}c^{4} $.

Note that if we put $ F_{\sigma}^{(-)} = - F_{\sigma}^{(+)} $ and $ G_{\sigma}^{(-)} = G_{\sigma}^{(+)} $ in the second system (or vice versa, $ F_{\sigma}^{(-)} = F_{\sigma}^{(+)} $ and $ G_{\sigma}^{(-)} = - G_{\sigma}^{(+)} $), it will coincide with the first one. Thus, we can consider only one system for functions $ F_{\sigma} $ and $ G_{\sigma} $ using  the  relations 
\begin{equation}
\label{F,G_p,m} 
F_{\sigma}^{(+)} = F_{\sigma} , \quad F_{\sigma}^{(-)} = \sigma F_{\sigma} , \qquad G_{\sigma}^{(+)} = G_{\sigma} , \quad G_{\sigma}^{(-)} = - \sigma G_{\sigma} .
\end{equation}
Thus,  equations for the spin states $ \sigma = +1 $ and $ \sigma = -1 $ are 
\begin{equation}
\label{systFi_+} 
\text{for} \quad \sigma = +1 \quad \begin{array}{c}
-ic\hbar \frac{dF_{+}}{dr_{\perp}} + i\frac{c\hbar M}{r_{\perp}} F_{+} - \left( \mathcal{E}_{k} + E \right) G_{+} + V_{0}(r_{\perp}) G_{+} = 0 , \\ 
-ic\hbar \frac{dG_{+}}{dr_{\perp}} - i\frac{c\hbar M}{r_{\perp}} G_{+} + \left( \mathcal{E}_{k} - E \right) F_{+} + V_{0}(r_{\perp}) F_{+} = 0 ,
\end{array}  
\end{equation}
and 
\begin{equation}
\label{systFi_-} 
\text{for} \quad \sigma = -1 \quad  \begin{array}{c}
-ic\hbar \frac{dF_{-}}{dr_{\perp}} + i\frac{c\hbar M}{r_{\perp}} F_{-} + \left( \mathcal{E}_{k} - E \right) G_{-} + V_{0}(r_{\perp}) G_{-} = 0 , \\ 
-ic\hbar \frac{dG_{-}}{dr_{\perp}} - i\frac{c\hbar M}{r_{\perp}} G_{-} - \left( \mathcal{E}_{k} + E \right) F_{-} + V_{0}(r_{\perp}) F_{-} = 0 ,
\end{array}
\end{equation}

At large distances $ r_{\perp} \rightarrow \infty $ the potential $ V_{0} \left( r_{\perp} \right) \rightarrow 0 $ as well as the term $ 1/r_{\perp} \rightarrow 0 $, and the  asymptotic behavior of functions are determined by the following equations 
\[
ic\hbar \frac{dF_{\sigma}}{dr_{\perp}} + \left( E + \sigma \mathcal{E}_{k} \right) G_{\sigma} = 0 , \quad ic\hbar \frac{dG_{\sigma}}{dr_{\perp}} + \left( E - \sigma \mathcal{E}_{k} \right) F_{\sigma} = 0 , 
\] 
from which it follows that
\[
c^{2}\hbar^{2} \frac{d^{2}F_{\sigma}}{dr_{\perp}^{2}} = \left( \mathcal{E}_{k}^{2} - E^{2} \right) F_{\sigma} 
\]
(the same equation takes place for $ G_{\sigma} $). For the bound electron states which we will consider below, at $ r_{\perp} \rightarrow \infty $ the function behavior must be exponential,  $ F_{\sigma} \sim \exp (- \varkappa_{\perp} r_{\perp}) $ which takes place at $ E^{2} < \mathcal{E}_{k}^{2} = m^{2}c^{4} + c^{2} \hbar^{2} k^{2} $. Therefore, the in-plane damping
\begin{equation}
\label{varkappa} 
 \varkappa _{\perp}=  
\frac{1}{c\hbar} \sqrt{\mathcal{E}_{k}^{2} - E^{2} }   
\end{equation} 
is a real number. It shows that damping outside the chain depends on the  energy of the longitudinal electron motion. 

To find solution of Eq. (\ref{systFi_+}) for the spin state $ \sigma = +1 $ one can use the substitution 
\begin{equation}
\label{subsFG_+} 
F_{+} = \rho^{-1/2} e^{-\xi /2} f_{+}(r_{\perp}) , \quad G_{+} = i \rho^{1/2} e^{-\xi /2} g_{+}(r_{\perp}) , \quad \rho = \sqrt{\frac{\mathcal{E}_{k} - E}{\mathcal{E}_{k} + E}} = \frac{c\hbar \varkappa_{\perp}}{\mathcal{E}_{k} + E} , 
\end{equation}
in which the dimensionless variable is used
\begin{equation}
	\label{xi}
\xi = 2 \varkappa _{\perp} r_{\perp},
\end{equation}
which transforms the system to the following equations for functions $ f $ and $ g $
\[
\begin{array}{c}
 \frac{df_{+}}{dr_{\perp}} - \varkappa_{\perp} f_{+} - \frac{M}{r_{\perp}} f_{+} + \varkappa_{\perp} g_{+} - \frac{\mathcal{E}_{k}}{c^{2}\hbar^{2}\varkappa}_{\perp} V_{0}(r_{\perp}) g_{+} + \frac{E}{c^{2}\hbar^{2}\varkappa_{\perp}} V_{0}(r_{\perp}) g_{+} = 0 , \\ 
 \frac{dg_{+}}{dr_{\perp}} - \varkappa_{\perp} g_{+} + \frac{M}{r_{\perp}} g_{+} + \varkappa_{\perp} f_{+} + \frac{\mathcal{E}_{k}}{c^{2}\hbar^{2}\varkappa} V_{0}(r_{\perp}) f_{+} + \frac{E}{c^{2}\hbar^{2}\varkappa} V_{0}(r_{\perp}) f_{+} = 0 . 
\end{array}
\] 

The atomic chain potential $ V_{0} \left( r_{\perp} \right) $  is given in Eq. (\ref{V_0}) where we write down the atom charge $ Q $ as $ Q = e Z_{v} $ with $ Z_{v} $ being the number of the valence electrons in an atom. At $ r_{\perp} > a $, the potential can be expanded in the series 
\[
V_{0} \left( r_{\perp} \right) = - \frac{2e^{2}Z_{v}}{a} \ln \left( 1 + \frac{a}{2r_{\perp}} + \frac{a^{2}}{8r_{\perp}^{2}} + \ldots \right) \approx - \frac{e^{2}Z_{v}}{r_{\perp}} \left( 1 - \frac{a^{2}}{24r_{\perp}^{2}} + \ldots \right) ,
\]
and, to obtain the solution for the electrons bound by the chain, we can use the  effective potential $ V_{eff} = C_{eff}e^{2}Z_{v}/r_{\perp} $ with the parameter $ C_{eff} $. 

By adding and subtracting equations for functions $ f $ and $ g $, we obtain the system of equations for functions $ u_{+}^{(\pm)} = f_{+} \pm g_{+} $
\[ 
\begin{array}{c} 
\frac{du_{+}^{(+)}}{dr_{\perp}} - \frac{E}{c^{2}\hbar^{2}\varkappa_{\perp}} \frac{eQ}{\mathbf{r}_{\perp}} u_{+}^{(+)} - \frac{M}{r_{\perp}} u_{+}^{(-)} - \frac{\mathcal{E}_{k}}{c^{2}\hbar^{2}\varkappa_{\perp}} \frac{eQ}{r_{\perp}}  u_{+}^{(-)} = 0 , \\ 
\frac{du_{+}^{(-)}}{dr_{\perp}} - 2\varkappa_{\perp} u_{+}^{(-)} + \frac{E}{c^{2}\hbar^{2}\varkappa_{\perp}} \frac{eQ}{r_{\perp}} u_{+}^{(-)} - \frac{M}{r_{\perp}} u_{+}^{(+)} + \frac{\mathcal{E}_{k}}{c^{2}\hbar^{2}\varkappa_{\perp}} \frac{eQ}{r_{\perp}} u_{+}^{(+)} = 0 .
\end{array} 
\]

Using the substitution $ u_{+}^{(+)} = r_{\perp}^{\gamma} p(r_{\perp}) $ and $ u_{+}^{(-)} = r_{\perp}^{\gamma} q(r_{\perp}) $, we obtain the system of equations for functions $ p(r_{\perp}) $ and $ q(r_{\perp}) $ 
\begin{equation}
\label{syst-p,q} 
\begin{array}{c}
r_{\perp} \frac{dp(r_{\perp})}{dr_{\perp}} + \left( \gamma - \frac{E C_{eff}eQ}{c^{2}\hbar^{2}\varkappa_{\perp}} \right) p (r_{\perp})= \left( M + \frac{\mathcal{E}_{k}C_{eff}eQ}{c^{2}\hbar^{2}\varkappa_{\perp}} \right) q (r_{\perp}), \\ 
r_{\perp} \frac{dq(r_{\perp})}{dr_{\perp}} + \left( \gamma + \frac{EC_{eff}eQ}{c^{2}\hbar^{2}\varkappa_{\perp}}  - 2\varkappa_{\perp} r_{\perp} \right) q(r_{\perp}) = \left( M - \frac{\mathcal{E}_{k}C_{eff}eQ}{c^{2}\hbar^{2}\varkappa_{\perp}} \right) p (r_{\perp}).
\end{array}
\end{equation}

This system of two first order equations are reduced to equations of the second order for each functions  
\[
r_{\perp}^{2} \frac{d^{2}p(r_{\perp})}{dr_{\perp}^{2}} + r_{\perp} \left( 1 + 2\gamma - 2 \varkappa_{\perp} r_{\perp} \right) \frac{dp(r_{\perp})}{dr_{\perp}} + 
\]
\[
\quad +\left[ \gamma^{2} - M^{2} + C_{eff}^{2}Z_{v}^{2}\alpha^{2} + 2 \varkappa_{\perp} r_{\perp} \left( \frac{EC_{eff}eQ}{c^{2}\hbar^{2}\varkappa_{\perp}} - \gamma \right) \right] p (r_{\perp})= 0 , 
\] 
\[
r_{\perp}^{2} \frac{d^{2}q(r_{\perp})}{dr_{\perp}^{2}} + r_{\perp} \left( 1 + 2\gamma - 2\varkappa_{\perp} r_{\perp} \right) \frac{dq(r_{\perp})}{dr_{\perp}} +
\]
\[\quad + \left[ \gamma^{2} - M^{2} + C_{eff}^{2}Z_{v}^{2}\alpha^{2} +  2\varkappa_{\perp} r_{\perp} \left( \frac{EC_{eff}eQ}{c^{2}\hbar^{2}\varkappa_{\perp}} - \gamma - 1 \right) \right] q(r_{\perp}) = 0 .
\] 
Here $ e^{2} /\hbar c = \alpha $ is the Sommerfeld fine structure constant and we have used the definition (\ref{varkappa}).
 
We see that functions $ p (r_{\perp})$ and $ q(r_{\perp}) $ satisfy the hypergeometric differential equation. The solution is to be finite throughout the space and, therefore, the  hypergeometric series should terminate for some value of $ n $. This condition leads to the equalities 
\begin{equation}
\label{cond-gamma,E} 
\gamma^2 \equiv \gamma _{M}^{2} = M^{2} - C_{eff}^{2}Z_{v}^{2}\alpha^{2} , \quad \text{and} \quad 
\frac{EC_{eff}eQ}{c^{2}\hbar^{2}\varkappa_{\perp}} - \gamma _{M} - 1 = n 
\end{equation}
where $ n $ is a positive integer number, $ n = 0,1,2,\ldots $. Under this condition, the functions $ p(r_{\perp}) $ and $ q(r_{\perp}) $ satisfy the following differential equations  
\begin{equation}
\label{Eqs-p,q} 
\begin{array}{c}
\xi \frac{d^{2}p(\xi)}{d\xi ^{2}} + \left( 1 + 2\gamma_M - \xi \right) \frac{dp(x)}{d\xi} + \left( n + 1 \right) p(\xi) = 0 , \\ 
\xi \frac{d^{2}q(\xi )}{d\xi ^{2}} + \left( 1 + 2\gamma_M - \xi \right) \frac{dq(\xi )}{d\xi} + n q(\xi) = 0 ,
\end{array} 
\end{equation}
for generalized Laguerre polynomial $ \mathit{L}_{n}^{2\gamma_M} \left( \xi \right) $ of the variable $ \xi $ determined in Eq. (\ref{xi}).  

Therefore, we obtain the solution $ p(r_{\perp}) = C_{1} \mathit{L}_{n+1}^{2\gamma_M} \left( 2\varkappa_{\perp} r_{\perp} \right) $ and $ q(r_{\perp}) = C_{2} \mathit{L}_{n}^{2\gamma_M} \left( \xi  \right) $ with 
\begin{equation}
\label{gamma} 
\gamma_M = \sqrt{M^{2} - C_{eff}^{2} Z_{v}^{2} \alpha^{2} }
\end{equation}
according to the first equality in the system (\ref{cond-gamma,E}). Here the square root has the positive sign, only, as it follows from the condition of the normalization integral convergence. The second equality in (\ref{cond-gamma,E}) gives the energy eigenvalue 
\begin{equation}
\label{E} 
E = E_{k,M,n} = \mathcal{E}_{k} \Delta_{n,M} , 
\end{equation} 
where 
\begin{equation}
	\label{Delta} 
\Delta_{n,M} = \frac{\gamma_M + n}{\sqrt{\left(\gamma_M + n \right)^{2} + C_{eff}^{2} Z_{v}^{2} \alpha^{2}}} = \sqrt{1 - \frac{C_{eff}^{2} Z_{v}^{2} \alpha^{2}}{n^{2} + M^{2} + 2n\gamma_M}} ,
\end{equation} 
$n=1,2, \dots $ and $ \mathcal{E}_{k} $ is determined in Eq. (\ref{calE_k}). The damping parameter (\ref{varkappa}) takes the value 
\begin{equation}
\label{kappa-E} 
\varkappa_{\perp} = \frac{1}{r_{B}} \frac{C_{eff}Z_{v}}{\sqrt{n^{2} + M^{2} + 2n\gamma_M }} \sqrt{1 + \alpha^{2}r_{B}^{2} k^{2}} 
\end{equation} 
where $ r_{B} = \hbar^{2} /me^{2} $ is the Bohr radius, and parameter $ \rho $ in Eq. (\ref{subsFG_+}) is 
\begin{equation}
\label{beta} 
\rho = \sqrt{\frac{\mathcal{E}_{k} - E_{n,k}}{\mathcal{E}_{k} + E_{n,k}}} = 
\sqrt{\frac{1 - \Delta_{n,M}}{1 + \Delta_{n,M}}} = \frac{C_{eff} Z_{v} \alpha}{\left( 1 + \Delta_{n,M} \right)\sqrt{n^{2} + M^{2} + 2n\gamma_M} } .
\end{equation}

The condition that functions  $ p(r_{\perp}) $ and $ q(r_{\perp})  $ are the solution of the system (\ref{syst-p,q}), determines the relation between constants $ C_{1} $ and $ C_{2} $ (cf. Eq. (\ref{beta})): 
\begin{equation}
	\label{c1-c2} 
 \left( 2\gamma_M + n + 1 \right) C_{1}^{(+)} = - \left(\delta + M \right) C_{2}^{(+)}  , \quad 
 \left( n + 1 \right) C_{2}^{(+)} = - \left( \delta - M \right) C_{1}^{(+)}  
\end{equation}
where 
\begin{equation}
\label{b} 
\delta = \sqrt{M^{2} + (1+n)(2\gamma_M + 1 + n)} . 
\end{equation}

Let us now consider the spin states with $ \sigma = -1 $. They are described by Eqs. (\ref{systFi_-}) from which by substitution 
\begin{equation}
\label{subsF,G_-} 
F_{-} = i \rho^{1/2} e^{-\xi /2} r_{\perp}^{\gamma_M} \frac{1}{2} \left( p(r_{\perp}) + q(r_{\perp}) \right)  , \quad G_{-} = \rho ^{-1/2} e^{-\xi /2} r_{\perp}^{\gamma_M} \frac{1}{2} \left( p(r_{\perp}) - q(r_{\perp}) \right) 
\end{equation}
we obtain the equations for functions $ p(r_\perp) $ and $ q (r_\perp) $:
\begin{equation}
\label{syst-p,q(-)} 
\begin{array}{c}
r_{\perp} \frac{dp_{-}(r_\perp) }{dr_{\perp}} + \left( \gamma_M - \frac{E C_{eff}eQ}{c^{2}\hbar^{2}\varkappa_{\perp}} \right) p_{-}(r_\perp)  + \left( M - \frac{\mathcal{E}_{k}C_{eff}eQ}{c^{2}\hbar^{2}\varkappa_{\perp}} \right) q_{-}(r_\perp)  = 0 , \\ 
r_{\perp} \frac{dq_{-}(r_\perp) }{dr_{\perp}} + \left( \gamma_M + \frac{EC_{eff}eQ}{c^{2}\hbar^{2}\varkappa_{\perp}}  - 2\varkappa_{\perp} r_{\perp} \right) q_{-}(r_\perp)  + \left( M + \frac{\mathcal{E}_{k}C_{eff}eQ}{c^{2}\hbar^{2}\varkappa_{\perp}} \right) p_{-} (r_\perp) = 0 .
\end{array}
\end{equation}
from which it follows that under conditions (\ref{cond-gamma,E}) $ p(r_\perp) $ and $ q(r_\perp)$ are the solutions of Eqs. (\ref{Eqs-p,q}) and, therefore, $ p(r_{\perp}) = C_{1} \mathit{L}_{n+1}^{2\gamma_M} \left( 2\varkappa_{\perp} r_{\perp} \right) $ and $ q(r_{\perp}) = C_{2} \mathit{L}_{n}^{2\gamma_M} \left( 2\varkappa_{\perp} r_{\perp} \right) $. In this case $ p(r_\perp) $ and $ q(r_\perp) $ ought to be the solutions of Eqs.(\ref{syst-p,q(-)}) which gives us the relation between constants $ C_{1} $ and $ C_{2} $ (cf. Eq. (\ref{c1-c2})):  
\begin{equation}
	\label{c1-c2-2} 
 \left( 2\gamma_M + n + 1 \right) C_{1}^{(-)} = - \left( b - M \right) C_{2}^{(-)}  , \quad 
\left( n + 1 \right) C_{2}^{(-)} = - \left( b + M \right) C_{1}^{(-)}  ,
\end{equation} 
Here constants $ C_{j}^{(\pm)} $ are determined from the normalization condition 
\[
\int \Psi^{\dagger}_{n,k,M,\sigma} \Psi_{n,k,M,\sigma} d^{3}r = \int_{0}^{\infty} r_{\perp} dr_{\perp} \int_{0}^{2\pi} d\varphi \int_{-L/2}^{L/2} dz \Psi^{\dagger}_{n,k,M,\sigma} \Psi_{n,k,M,\sigma} = 1 
\]
with account of the relation between constants $ C_{1} $ and $ C_{2} $. 

As a result, we have orthonormal eigen bispinors for the spin state $ \sigma $   
\begin{equation}
\label{Psi_+} 
\Psi_{n,k,M,+} =  A_{n,k,M} e^{i \left( kz + M\varphi \right) }e^{-\xi /2} r_{\perp}^{\gamma_M - 1/2} \left( 
\begin{array}{c}
 \rho^{-1/2} P_{n,+}\left( \xi  \right) \chi_{\uparrow} \\
 i \rho^{1/2} Q_{n,+}\left( \xi \right) \chi_{\downarrow}
\end{array} \right)  
\end{equation} 
and 
\begin{equation}
\label{Psi_-} 
\Psi_{n,k,M,-} =  A_{n,k,M} e^{i \left( kz + M\varphi \right) }e^{-\xi/2} r_{\perp}^{\gamma_M - 1/2} \left( 
\begin{array}{c}
\rho^{-1/2} P_{n,-}\left( \xi  \right) \chi_{\downarrow} \\
i \rho^{1/2} Q_{n,-}\left( \xi  \right) \chi_{\uparrow}
\end{array} \right) 
\end{equation} 
where the spinors $ \chi_{\uparrow /\downarrow} $ are given by expressions 
\begin{equation}
	\label{xi-ud} 
\chi_{\uparrow} = \left( \begin{array}{c}
e^{-i\varphi /2} \\
0
\end{array}
 \right) , \quad \chi_{\downarrow} = \left( \begin{array}{c}
0 \\
e^{i\varphi /2}
\end{array}
 \right) ,
\end{equation} 
$ A_{n,k,M} $ is the normalization constant 
\begin{equation}
	\label{Ankm} 
A_{n,k,M} = \sqrt{\frac{c\hbar \left( 2\varkappa_{\perp} \right)^{2\gamma_M + 2} }{4\pi L b \mathcal{E}_{k}}} ,
\end{equation} 
and $ P_{n,\sigma} $ and $ Q_{n,\sigma} $ are the following polynomials 
\[
\begin{array}{c}
P_{n,+}\left( 2\varkappa _{\perp}\rho \right) = \sqrt{b + M} \tilde{\mathit{L}}_{n+1}^{2\gamma} \left( 2\varkappa_{\perp} \rho \right) - \sqrt{b - M} \tilde{\mathit{L}}_{n}^{2\gamma} \left( 2\varkappa _{\perp}\rho \right) , \\ 
Q_{n,+}\left( 2\varkappa _{\perp}\rho \right) = \sqrt{b + M} \tilde{\mathit{L}}_{n+1}^{2\gamma} \left( 2\varkappa _{\perp}\rho \right) + \sqrt{b - M} \tilde{\mathit{L}}_{n}^{2\gamma} \left( 2\varkappa_{\perp} \rho \right) , \\ 
P_{n,-}\left( 2\varkappa _{\perp}\rho \right) = \sqrt{b - M} \tilde{\mathit{L}}_{n+1}^{2\gamma} \left( 2\varkappa _{\perp}\rho \right) - \sqrt{b + M} \tilde{\mathit{L}}_{n}^{2\gamma} \left( 2\varkappa _{\perp}\rho \right) , \\ 
Q_{n,-}\left( 2\varkappa _{\perp}\rho \right) = \sqrt{b - M} \tilde{\mathit{L}}_{n+1}^{2\gamma} \left( 2\varkappa _{\perp}\rho \right) + \sqrt{b + M} \tilde{\mathit{L}}_{n}^{2\gamma} \left( 2\varkappa _{\perp}\rho \right) . 
\end{array} 
\] 
in which functions $ \tilde{\mathit{L}}_{n}^{2\gamma_M} $ are the normalized Laguerre polynomials
\begin{equation}
\label{lag}
\tilde{\mathit{L}}_{n}^{2\gamma_M} \left(\xi \right) = \sqrt{\frac{ n!}{\Gamma \left( 2\gamma_M + n + 1 \right)}} \mathit{L}_{n}^{2\gamma_M}\left( \xi \right) .
\end{equation} 

Note that states with the same quantum numbers $ n,k,M $ are degenerate with respect to spin index $ \sigma $. This degeneracy is accidental because besides the integral of motion $  \hat{\mathcal{S}}_{z} $ the DE with the field $ Q/r_{\perp} $ admits one more integral of motion, namely, 
\begin{equation}
\label{newInv} 
\hat{A}_{BEL} = \bm{\hat{\Omega}} \cdot \bm{\hat{\Pi}}_{\perp} + \frac{eQ}{cr_{\perp}} \left( \bm{\hat{\Gamma}} \cdot \mathbf{r}_{\perp} \times \mathbf{e}_{z} \hat{p}_{z} + \hat{\Gamma}_{z} \hat{L}_{z} + \frac{\hbar}{2} \hat{\rho}_{2} - mc \bm{\hat{\Sigma}} \cdot \mathbf{r}_{\perp} \right)   
\end{equation} 
in which  
\[
\bm{\hat{\Pi}}_{\perp} = \hat{\Pi}_{x} \mathbf{e}_{x} + \hat{\Pi}_{y} \mathbf{e}_{y} , \quad \hat{\Pi}_{x} = \frac{1}{2} \left( \hat{p}_{y} \hat{L}_{z} + \hat{L}_{z} \hat{p}_{y} \right) , \quad \hat{\Pi}_{y} = - \frac{1}{2} \left( \hat{p}_{x} \hat{L}_{z} + \hat{L}_{z} \hat{p}_{x} \right) 
\] 
(for other notations see Appendix).
Existence of this new invariant establishes the "accidental" degeneracy in the field $ \sim 1/r_{\perp} $ just as the existence of the Johnson-Lippmann invariant $\hat{A}_{JL} $ \cite{Johns-Lip} in the relativist Kepler problem (DE with the Coulomb potential) explains degeneracy of hydrogen-like spectrum with respect to the spinor quantum number (see also \cite{BEL2}).

\section{Conclusions}

Thus, within the Dirac description the exact analytical expressions for the three-dimensional bound electron states in the Coulomb field of the chain consisting  of positively charged ions, are obtained  using the new spinor invariant (\ref{newInv}) found for this problem. These solutions are given by the expressions (\ref{Psi_+})--(\ref{lag}) which prove existence of the  entanglement of different solutions.

It is worth to add that the problem of finding solutions of the DE is actively discussed in the literature for many years \cite{Arda,Yahya,Karayer}. For this artificial (non-physical) potentials are considered. On the contrary, in the present paper the exact analytical solution of the DE with the realistic Coulomb potential (\ref{V_0}), describing interaction of an electron with a chain of positively charged ions, is obtained.

The next step in this problem is to take into account  periodicity of a  chain potential and  interaction of an electron also with the deformation of a chain caused by the displacement of atoms from their equilibrium positions (cf. Davydov's soliton), using the obtained here solution and the new invariant (\ref{newInv}). 

\section{Appendix}\label{ApDM}

For convenience of the reader here we summarize Dirac matrices and their notations.
In the standard representation Dirac matrices have the form:
\begin{equation}
	\label{DMst}
	\hat{\beta} = \left( \begin{array}{cc}
		\hat{\mathbb{I}}_{2} & 0 \\ 0 & -\hat{\mathbb{I}}_{2}
	\end{array}  \right) \,, \quad \hat{\alpha}_{j} = \left( 
	\begin{array}{cc}
		0 & \hat{\sigma}_{j} \\ \hat{\sigma}_{j} & 0
	\end{array} \right) \, ,
\end{equation}
where $ \hat{\mathbb{I}}_{2} $ is a unit $ 2\times 2 $ matrix and $ \hat{\sigma}_{j} $ are Pauli matrices ($ j=x,y,z $). 

 Using products of two and more matrices from the above four ones, one can construct sixteen linearly independent matrices including a unit matrix. Any arbitrary  $ 4 \times 4 $ matrix can be  represented as a linear combination of matrices from this set. Physical observables correspond to the Hermitian operators, and, respectively, it is convenient to use Hermitian Dirac matrices 
\begin{equation}
	\label{product} 
	\begin{array}{c}
		\hat{\Gamma}_{j} = -i \hat{\beta} \hat{\alpha}_{j}  , \quad 
		\hat{\Sigma}_{j} = -i \hat{\alpha}_{k} \hat{\alpha}_{l} ,  \quad
		\hat{\Omega}_{j} = -i \hat{\beta} \hat{\alpha}_{k} \hat{\alpha}_{l} ,  \\ 
		\hat{\rho}_{1} = -i \hat{\alpha}_{j} \hat{\alpha}_{k} \hat{\alpha}_{l}  , \quad 
		\hat{\rho}_{2} = - \hat{\beta} \hat{\alpha}_{j} \hat{\alpha}_{k} \hat{\alpha}_{l}  ,
	\end{array}
\end{equation}
where indices $ j,k,l $ take values $ x $, $ y $, $ z $, three numbers  $ j,k,l $ are any cyclic  permutation of $ x,y,z $, and coefficient $ i $ is introduced in order to have the new matrices, denoted as  $ \hat{\Gamma}_{j} $, $ \hat{\Sigma}_{j} $, $ \hat{\Omega}_{j} $, $ \hat{\rho}_{1} $ and $ \hat{\rho}_{2} $, Hermitian, too. Using often used notation $ \hat{\rho}_{3} \equiv \hat{\beta} $ and adding a unitary matrix $ \hat{\mathbb{I}}_{4} $, we get sixteen linearly independent matrices
\begin{equation}
	\label{16m} 
	\hat{\mathbb{I}}_{4} , \quad \hat{\rho}_{1} , \quad \hat{\rho}_{2} , \quad \hat{\rho}_{3} , \quad \bm{\hat{\Sigma}} , \quad  \bm{\hat{\alpha}} , \quad \bm{\hat{\Gamma}} , \quad \bm{\hat{\Omega}} .  
\end{equation}
Here vector-matrices are introduced $ \bm{\hat{\Sigma}} $, $ \bm{\hat{\Gamma}} $ and $ \bm{\hat{\Omega}} $ which have $ x $, $ y $ and $ z $ spatial components, similar to the matrix  $ \bm{\hat{\alpha}} $. The set of linearly independent matrices (\ref{16m}) in the standard representation has the form
\begin{eqnarray}
	\label{DMs_st}
	\hat{\mathbb{I}}_{4} = \left( \begin{array}{cc}
		\hat{\mathbb{I}}_{2} & 0 \\ 0 & \hat{\mathbb{I}}_{2}
	\end{array}  \right) ,\: 
	\hat{\rho }_{1} = \left( \begin{array}{cc}
		0 & \hat{\mathbb{I}}_{2} \\ \hat{\mathbb{I}}_{2} & 0
	\end{array}  \right) ,\: 
	\hat{\rho}_{2} = \left( \begin{array}{cc}
		0 & -i \hat{\mathbb{I}}_{2} \\ i \hat{\mathbb{I}}_{2} & 0
	\end{array}  \right) ,\: 
	\hat{\rho}_{3} = \left( \begin{array}{cc}
		\hat{\mathbb{I}}_{2} & 0 \\ 0 & -\hat{\mathbb{I}}_{2}
	\end{array}  \right) , \nonumber \\ 
	\bm{\hat{\Sigma}} = \left( 
	\begin{array}{cc}
		\bm{\hat{\sigma}} & 0 \\ 0 & \bm{\hat{\sigma}}
	\end{array} \right) , \: 
	\bm{\hat{\alpha}} = \left( 
	\begin{array}{cc}
		0 & \bm{\hat{\sigma}} \\ \bm{\hat{\sigma}} & 0
	\end{array} \right) ,\: 
	\bm{\hat{\Gamma}} = \left( 
	\begin{array}{cc}
		0 & -i\bm{\hat{\sigma}} \\ i\bm{\hat{\sigma}} & 0
	\end{array} \right) ,\: 
	\bm{\hat{\Omega}} = \left( 
	\begin{array}{cc}
		\bm{\hat{\sigma}} & 0 \\ 0 & -\bm{\hat{\sigma}}
	\end{array} \right) . 
\end{eqnarray}

\vskip5mm 
{\bf Acknowledgement.} 
\textit{This work was supported by the fundamental scientific program 0122U000887 of the Department of Physics and Astronomy of the National Academy of Sciences of Ukraine. The authors acknowledge support from the Simons Foundation (NY).}


\begin{thebibliography}{99} 
\bibitem{DauPeyr}
T. Dauxois, M.  Peyrard.  Physics of Solitons. Cambridge, Cambridge University Press, 2006.
\bibitem{KosIvKov}
A.M. Kosevich, B.A. Ivanov, and A.S. Kovalev. Magnetic solitons. Phys. Rep. 194, No 3/4, 117-238 (1990).
\bibitem{DavydovKislukha1}
A.S. Davydov, N.I. Kislukha. Solitary excitons in one-dimensional molecular chains. Phys. Stat. Sol. B (1973) \textbf{5} 465.  https://doi.org/10.1002/pssb.2220590212
\bibitem{Davydov}
A.S. Davydov. Solitons in Molecular Systems. Dordrecht: Reidel; 1991.
\bibitem{Scott}
A.C. Scott. Davydov's soliton. Phys. Rep. (1992) \textbf{217} 1.
\bibitem{KovalevKosevich}
A.M. Kosevich, A.S. Kovalev. Autolocalization of vibrations in one-dimensional anharmonic chain. (In Russian). Zh. Exp. Theor. Phys., (1974) \textbf{67} 1795 
\bibitem{DavQBiol}
A.S. Davydov. Biology and Quantum Mechanics. Pergamon Press, 1982. - Science - 229 pages. ISBN-10:0080263925; ISBN-13: 008026392.2
\bibitem{ground}
L.S. Brizhik, A.A. Eremko. Ground state diagram of a 1D electron-phonon system. Synth. Met., 2000, v. 109, No 1-3, 117-121. 
\bibitem{Brizhik-DAA}
L.S. Brizhik, J. Luo, B.M.A.G.  Piette, W.J. Zakrzewski. Long-range electron transport mediated by alpha-helices. Phys. Rev. E (2019) \textbf{100} 062205.  DOI: 10.1103/PhysRevE.100.062205.
\bibitem{BEL1}
A.A. Eremko, L. Brizhik, V.M. Loktev. Algebra of the spinor invariants and the relativistic hydrogen atom. Annals of Physics, 2023, 451, 169266. https://doi.org/10.1016/j.aop.2023.169266; 
\bibitem{BEL2}
A.A. Eremko, L.S. Brizhik,V.M. Loktev, Spin relevant invariants and the general solution of the Dirac equation for the Coulomb field. Annals of Physics, 439 (2022) 168786 (20 pp). https://doi.org/10.1016/j.aop.2022.168786. 
\bibitem{Johns-Lip}
M. H. Johnson and B. A. Lippmann, Relativistic Kepler Problem,  Phys. Rev. \textbf{78} (1950) 329(A).
\bibitem{Arda}
Altu\v{g} Arda, and Ramazan Sever, "Bound-state solutions of the Dirac equation for the Kratzer potential with pseudoscalar-Coulomb term", Eur. Phys. J. Plus (2019) \textbf{134}: 29
DOI 10.1140/epjp/i2019-12421-9 . 
\bibitem{Yahya}
W. A. Yahya and K. J. Oyewumi, "Bound state solutions of the Dirac equation for the
trigonometric and hyperbolic Scarf-Grosche potentials using the Nikiforov-Uvarov method", 
J. Math. Phys. \textbf{54}, 013508 (2013); https://doi.org/10.1063/1.4772478 

\bibitem{Karayer}
H. Karayer, "Analytical solution of the Dirac equation for the hyperbolic potential by the extended Nikiforov-Uvarov method", Eur. Phys. J. Plus (2019) \textbf{134}: 452
DOI 10.1140/epjp/i2019-12828-2. 
	
\end{thebibliography}
\end{document}